\documentclass[showpacs,preprintnumbers,amsmath,amssymb,superscriptaddress,nofootinbib,english]{revtex4}

\usepackage{natbib}
\usepackage{graphicx}
\usepackage{dcolumn}
\usepackage{bm}
\usepackage{epsfig}
\usepackage{graphicx}
\usepackage{hyperref}
\usepackage[usenames]{color}
\usepackage{url}

\hypersetup{
    colorlinks=true,
    linkcolor=green,
    citecolor=blue,
}

\newcommand{\remove}[1]{}

\newcommand{\rr}{\mathrm}
\newcommand{\atrans}{{a_{\rr{trans}}}}
\newcommand{\ztrans}{{z_{\rr{trans}}}}

\newcommand{\Mpc}{\ \rr{Mpc}}

\newcommand{\fMG}{f_{\rr{MG}}}
\newcommand{\betab}{\beta_{\rr{b}}}

\newcommand{\deltab}{\delta_{\rr{b}}}

\newcommand{\deltagamma}{\delta_{\gamma}}
\newcommand{\thetab}{\theta_{\rr{b}}}

\newcommand{\thetagamma}{\theta_{\rr\gamma}}
\newcommand{\sigmagamma}{\sigma_{\rr\gamma}}
\newcommand{\tauc}{\tau_{\rr{c}}}

\newcommand{\deltaphi}{\delta \phi}

\def\be{\begin{equation}}
\def\ee{\end{equation}}
\def\ba{\begin{eqnarray}}
\def\ea{\end{eqnarray}}

\begin{document}

\title{Early Modified Gravity: Implications for Cosmology}

\author{Philippe~Brax}
\email[Email address: ]{philippe.brax@cea.fr}
\affiliation{Institut de Physique Theorique, CEA, IPhT, CNRS, URA 2306, F-91191Gif/Yvette Cedex, France}
\author{Carsten van de Bruck}
\email[Email address: ]{C.vandeBruck@sheffield.ac.uk}
\affiliation{Consortium for Fundamental Physics, School of Mathematics and Statistics, University of Sheffield, Hounsfield Road, Sheffield, S3 7RH, United Kingdom}
\author{Sebastien Clesse}
\email[Email address: ]{sebastien.clesse@unamur.be}
\affiliation{Namur Center of Complex Systems (naXys), Department of Mathematics, University of Namur, Rempart de la Vierge 8, 5000 Namur, Belgium}
\affiliation{Physik Department T70, Excellence Cluster Universe, James-Franck-Strasse,
Technische Universit\"at M\"unchen, 85748 Garching, Germany}
\author{Anne-Christine~Davis}
\email[Email address: ]{a.c.davis@damtp.cam.ac.uk}
\affiliation{DAMTP, Centre for Mathematical Sciences, University of Cambridge, Wilberforce Road, Cambridge CB3 0WA, UK}
\author{Gregory Sculthorpe}
\email[Email address: ]{app09gis@sheffield.ac.uk}
\affiliation{Consortium for Fundamental Physics, School of Mathematics and Statistics, University of Sheffield, Hounsfield Road, Sheffield, S3 7RH, United Kingdom}

\date{\today}

\begin{abstract}
We study the effects of modifications of gravity after Big Bang Nucleosynthesis (BBN) which would manifest themselves mainly before recombination. We consider their effects on the Cosmic Microwave Background (CMB) radiation
and on the formation of large scale structure. The models that we introduce here represent all screened modifications of General Relativity (GR) which evade the local tests of gravity such as the violation of the strong equivalence principle as constrained  by the Lunar Ranging experiment. We use the tomographic description of modified gravity which defines models with screening mechanisms of the chameleon or Damour-Polyakov types and allows one to relate the temporal evolution of the mass and the coupling to matter of a scalar field to its Lagrangian and also to cosmological perturbations. The models with early modifications of gravity all involve a coupling to matter which is stronger in the past leading to effects on perturbations before recombination while minimising deviations from $\Lambda$CDM structure formation at late times. We find that a new family of early transition models lead to discrepancies in the CMB spectrum which could reach a few percent and appear as both enhancements and reductions of power for different scales.

\end{abstract}

\maketitle

\section{Introduction}
The observed accelerated expansion of the Universe~\cite{Riess:1998cb} can be explained within General Relativity (GR) by including the cosmological constant $\Lambda$, a new parameter whose presence is not forbidden by the basic postulates or symmetries of GR. Allowing for $\Lambda$ and some form of cold dark matter (CDM), cosmologists have arrived at a standard model of cosmology, the $\Lambda$CDM model, whose predictions are in remarkable agreement with observations, such as those from the cosmic microwave background radiation (CMB) and large scale structures in the Universe. But both CDM and the cosmological constant require some deeper understanding. Within extensions of the standard model of particle physics, CDM may find an explanation such as new particles in models with supersymmetry~\cite{Ellis:2010kf}. The cosmological constant term, however, has yet to be embedded within field theories of particle physics.

Instead of a $\Lambda$-term in Einstein's theory, equivalent to a constant energy density which has to be fine-tuned, many cosmologists would prefer a dynamical explanation for the accelerated expansion~\cite{Copeland:2006wr}, preferably one which does not depend on some initial conditions. Therefore, models have been developed in which a dynamical field, usually a scalar field, drives the accelerated expansion of the universe at late times~\cite{Clifton:2011jh}.  If the field couples to matter (baryonic or not) and the forces are very long-ranged (typically of the order of the horizon scale), the coupling is constrained to be very small, already at times when the photons decouple from baryonic matter in the early universe. In addition, local experiments~\cite{Adelberger:2002ic,Pourhasan:2011sm,Jain:2012tn,Williams:2012nc,Williams:2004qba,Williams:2003wu,Williams:2005rv} constrain the coupling of a long-ranged scalar field to normal matter to be smaller than $10^{-5}$ that of gravity. Given the strong constraints on long-ranged forces, other types of field theories have been studied, in particular screened models of modified gravity~\cite{Khoury:2010xi}. In these models, the gravitational sector is modified to include a scalar degree of freedom, whose interactions with ambient matter make the field either short ranged (as in chameleon/$F(R)$ theories~\cite{Khoury:2003aq,Khoury:2003rn,Brax:2004qh,Mota:2006ed,Brookfield:2006mq,Faulkner:2006ub,Brax:2008hh, Appleby:2007vb,Hu:2007nk}) or the effective coupling becomes small in region of high density (as it is the case of symmetron~\cite{Hinterbichler:2010es,Brax:2011pk,Hinterbichler:2011ca,Pietroni:2005pv} and/or dilaton models~\cite{Brax:2010gi,Biswas:2004be}). These theories behave effectively as the $\Lambda$CDM model for the background evolution. The evolution of perturbations on scales outside the Compton wavelength of the scalar field is expected to differ only slightly from $\Lambda$CDM. Only on smaller scales, deviations from $\Lambda$CDM are expected in these models. The interaction range usually changes with the cosmological expansion in these models as can the coupling strength, but usually local constraints today impose that in the present universe the range of interaction is less than 1 Mpc. Therefore to test these models with cosmological observations, predictions for structure formation at small length scales need to be studied in detail using N-body simulations~\cite{Brax:2011ja,Brax:2013mua,Brax:2012nk,Davis:2011pj}. Larger scales affected by linear perturbation theory can show deviations which are smaller but could still be within reach by precision cosmology. In the future, besides the large scale structure surveys at low redshifts, the 21-cm signal from the reionisation or the late dark ages will possibly enable us to draw a tomographic view of the matter distribution up to redshifts $z \sim 25$~\cite{Pritchard:2011xb,Furlanetto:2009qk,Clesse:2012th}, higher redshifts being much more difficult to probe with Earth-based radiotelescopes. At those redshifts, short scales which would be non-linear in the very late time Universe appear in the linear regime. 21-cm tomography therefore should be promising to constrain deviations from the linear growth of perturbations in General Relativity, and  thus to constrain modified gravity~\cite{Brax:2012cr}.

In this paper, we address the following questions: i) How can screened gravity affect the evolution of linear perturbations before recombination? And ii) imposing the BBN and local test constraints on screened modified gravity theory, can we find new regimes/models for which those effects leave observable imprints on the CMB as well as the matter power spectrum?
These questions are considered for two classes of screened gravity models: generalised chameleons, to which the $f(R)$, dilaton and chameleon models belong, and a new phenomenological model where the coupling of the scalar field to matter undergoes a transition before the time of last scattering.  Notice that viable $f(R)$ models are  typical examples of the application of the chameleon mechanism \cite{Brax:2008hh}. In the absence of this screening behaviour in the presence of matter, $f(R)$ models would lead to an enhancement of Newton's law locally which would be ruled out experimentally.  For this purpose, we use the fact that models of screened modified gravity can be described by two functions of the scale-factor $a$~\cite{Brax:2011aw,Brax:2012gr}: the effective mass $m(a)$ of the scalar field and the couplings to matter $\beta(a)$. This parameterisation is equivalent to describing the modified gravity models using field dependent interaction potentials $V(\phi)$ and coupling to matter functions $A(\phi)$. The simplest models such as $f(R)$ gravity and chameleons use an exponential function for $A(\phi)$ corresponding to a constant coupling $\beta$ to matter. This is enough to guarantee of the screening effects of the scalar field thanks to the chameleon mechanism in these models \cite{Brax:2008hh,Khoury:2003aq}.
For dilaton \cite{Brax:2010gi} and symmetron models \cite{Hinterbichler:2010es} where the scalar field is screened in dense environments thanks to the Damour-Polyakov effect \cite{Damour:1994zq}, the function $A(\phi)$ is not exponential anymore and the coupling to matter $\beta(\phi)$ becomes field dependent. In the  tomographic approach, all  these  models are described within one framework via different $m(a)$ and $\beta(a)$ functions.
We shall see that there can be significant deviations from $\Lambda$CDM in the CMB spectrum for the transition models where the modifications of gravity are significant before recombination. These manifest themselves as both enhancements and reductions of power on different scales. It is possible to choose parameters such that these differences are at the percent level while preserving structure formation and evading local tests of gravity.
In contrast we find that, for generalised chameleon models with couplings that are stronger at earlier rather than later times, the constraint on the variation over time of particle masses is violated before we observe effects on the CMB. For these models it would seem that large scale structure is the best cosmological test. Therefore we conclude that the signatures of the generalised chameleons and the early transition models are sufficiently different to envisage the possibility of distinguishing different models of modified gravity using CMB data if they were ever to be observed.

The paper is organised as follows: In the next Section, we briefly summarise models of screened modified gravity, using the $m(a)$, $\beta(a)$ parametrisation. We will also impose local constraints on those theories. In Section 3 we write down the perturbation equations and derive an analytical solution for the baryon perturbation during tight coupling. In Section 4 we study the evolution of the perturbations numerically. We conclude in Section 5. Details of the numerical implementation are summarised in the appendix.

\section{Screened Modified Gravity}

\subsection{Chameleon and Damour Polyakov mechanisms}

Scalar fields coupled to matter and with a long interaction range compared to the size of the solar system may play a role on cosmological scales as suggested by the discovery of the acceleration of the Universe. This would be in conflict with with experimental tests on the existence of fifth forces in the solar system such as the one given by the Cassini probe \cite{Bertotti:2003rm}.
Fortunately, the scalar field can be  screened in dense environments and therefore evade gravitational tests. This is the case for models subject to the
 chameleon and  Damour-Polyakov mechanisms. In the following, we will use models of this type which could have consequences on the Cosmic Microwave Background.

We focus on  scalar tensor theories defined by the action
\be
S=\int \mathrm d^4 x \sqrt{-g^E}\left[\frac{R_E}{16\pi G_N} -\frac{(\partial\phi)^2}{2} -V(\phi)\right]+S_{\mathrm m} \left[\psi, A^2(\phi) g_{\mu\nu}^E\right],
\ee
where $A(\phi)$ is an arbitrary function, $V(\phi)$ is the scalar potential and matter fields couple to the Jordan frame  metric
\be\label{confrel}
 g_{\mu\nu}^J= A^2(\phi) g_{\mu\nu}^E,
\ee
where the superscript E refers to the Einstein frame and J to the Jordan frame.
In the presence of non-relativistic matter, the dynamics of the scalar field are dictated  by the effective potential
\be
V_{\rm eff}(\phi)= V(\phi) + \left[ A(\phi)-1 \right] \rho_{\mathrm m} ,
\ee
where $\rho_{\mathrm m}$ is the matter energy density.
When the effective potential has a matter dependent minimum $\phi_{\rm min}(\rho_{\mathrm m})$, and if the effective mass of the scalar field
\be
m^2(\rho_{\mathrm m})= \left.  \frac{\mathrm d^2 V_{\rm eff}(\phi)}{\mathrm d \phi^2} \right|_{\phi=\phi_{\rm min}(\rho_{\mathrm m})}
\ee
becomes large enough to reduce the range of the scalar interaction, the effects of the scalar field on matter are screened~\cite{Khoury:2010xi}. This happens in models such as chameleons where the potential is an inverse power law and the coupling to matter
\be
\beta(\rho_{\mathrm m})= m_{\rm Pl} \left. \frac{\mathrm d\ln A(\phi)}{\mathrm d\phi}\right|_{\phi=\phi_{\rm min}(\rho_m)}
\ee
evaluated at the minimum, is constant. On the contrary if the coupling is not a constant anymore but converges to zero when the density increases, gravity tests on the existence of fifth forces can be evaded too \cite{Brax:2010gi}. This is the Damour-Polyakov mechanism \cite{Damour:1994zq}.

Let us first present a large class of models where the chameleon mechanism is necessary and at play. Consistent $f(R)$ models are highly relevant modifications of gravity which can serve as templates for the chameleon mechanism. Indeed, they must be in agreement with local constraints coming from gravitational tests. A way of fulfilling these bounds is to employ the chameleon mechanism in these theories. The $f(R)$ models where the action is given by
\be
S=\int d^4x \sqrt{-g_J} \frac{f(R_J)}{16\pi G_N}
\ee
can be written in the Einstein frame using the scalar field defined implicitly by
\be
\frac{\mathrm d f(R)}{ \mathrm d R}=e^{-2\beta\phi/m_{\rm Pl}},
\ee
where $\beta=\frac{1}{\sqrt{6}}$ and
we have
\be
A^2(\phi)=e^{2\beta\phi/m_{\rm Pl}}.
\ee
The interaction potential is
\be
V(\phi)= \frac{m^2_{\rm Pl}}{2} \frac{Rf_R -f}{f_R^2}.
\ee
Naively, one would conclude that $\beta$ is so large that these models are excluded by gravity tests. In fact, this is not case when the choice of $f(R)$ is such that $V(\phi)$ is a decreasing function $\phi$
and therefore $V_{\rm eff}$ admits a matter dependent minimum \cite{Brax:2008hh}. The chameleon mechanism applies for a wide range of models, such as \cite{Appleby:2007vb,Hu:2007nk}
\begin{eqnarray}
f(R) &=& \frac{R}{2} + \frac{1}{2a}\log \left[\cosh(aR)-\tanh(b)\sinh(aR)\right]~,\nonumber\\
f(R) &=& R - m^2 \frac{c_1 (R/m^2)^n}{c2(R/m^2)^n + 1}~.\nonumber
\end{eqnarray}
These models are excellent templates to study modified gravity in the late time Universe. In particular, in the limit of large curvature, the second model behaves as
\be\label{lcm}
f(R)=\Lambda + R -\frac{f_{R_0}}{n} \frac{R_0^{n+1}}{R^n}~.
\ee
We will later relate this model to a screened scalar field model via a tomographic map discussed below.

The string dilaton \cite{Damour:2002mi,Damour:2002nv} at strong coupling is an example where the Damour-Polyakov is at play,
\be
V(\phi)=V_0 e^{-\phi/m_{\rm Pl}},\ \ A(\phi)= 1+\frac{A_2}{2m_{\rm Pl}^2} (\phi-\phi_\star)^2
\ee
and $A_2>0$ to guarantee that the field converges to $\phi_{\star}$ in dense environments where the coupling to matter vanishes.
Another family of such models are the symmetrons \cite{Hinterbichler:2010es} where $V(\phi)$ is a symmetry breaking potential where the symmetry is restored at the origin and broken for $\phi=\phi_{\rm sym}$. When the function
\be
A(\phi)= 1+\frac{A_2}{2m_{\rm Pl}^2} \phi^2
\ee
is chosen, with $A_2>0$,  the symmetry is restored in dense environments and broken in vacuum.
All in all, we see that a variety of screened modified gravity models have been described using various functional forms for $V(\phi)$ and $A(\phi)$ which corresponds to field dependent couplings to matter
\be\label{betadef}
\beta(\phi)= m_{\rm Pl} \frac{\mathrm d\ln A(\phi)}{\mathrm d\phi}.
\ee
In fact a tomographic method has been devised where all these models can be captured by the sole time dependence of the coupling function
$\beta(a)$ and the mass $m(a)$ at the cosmological minimum of the effective potential as a function of the cosmological scale factor $a$~\cite{Brax:2011aw,Brax:2012gr}.
We will use this versatile approach explicitly in the following.

\subsection{Einstein vs Jordan Frames}
Before moving on to the description of the effects of early modifications of gravity on the CMB, we briefly comment on the Einstein frame (used in this paper) and its relation to the Jordan frame. When studying the growth of structure or the CMB, one needs to describe the cosmological perturbations of the models (for a thorough description of the cosmological perturbation theory in these models, see \cite{BraxDavisMGCMB}). This could be equivalently achieved using the Jordan or the Einstein frame. For studies of structure formation in modified gravity theories, working effectively in the Jordan frame, see \cite{Zhao,Bean} and references therein.

In the Einstein frame, the Einstein equation is not modified and for a single scalar field, no anisotropic stress is generated by the scalar field. This follows immediately from the Einstein equations. The metric perturbations can be described in the conformal Newtonian gauge where the metric reads
\begin{equation}
\mathrm ds_E^2= a_E^2(\eta) \left[-(1+2\Psi_E) \mathrm d\eta^2 + (1-2\Phi_E) \mathrm dx^2\right].
\end{equation}
In the case of vanishing anisotropic stress, Einstein's equations give $\Psi_E = \Phi_E$, which is valid in the Einstein frame.
The Jordan frame metric is related to the Einstein frame metric via a conformal transformation as in Eq. (\ref{confrel}).

Writing
\begin{equation}
\mathrm ds_J^2= a_J^2(\eta) \left[-(1+2\Psi_J) \mathrm d \eta^2 + (1-2\Phi_J) \mathrm d x^2 \right],
\end{equation}
one can easily show by perturbing (\ref{confrel}) that (see also \cite{Catena, Chiba})
\begin{eqnarray}
\Psi_J = \Psi_E + \frac{\mathrm d\ln A}{\mathrm d\phi}\delta \phi, ~~~\Phi_J = \Phi_E - \frac{\mathrm d\ln A}{\mathrm d\phi}\delta \phi.
\end{eqnarray}
Thus, if for example in the Einstein frame the anisotropic stress is zero, then $\Phi_E = \Psi_E$, but in the Jordan frame $\Psi_J - \Phi_J = 2\beta \frac{\delta\phi}{m_{\rm Pl}}$, with $\beta$ defined in (\ref{betadef}). We emphasise again that both frames are on the same footing (at least classically) and the calculations can be done in either frames.

In the following, we will consider the cosmological perturbations of the models. A convenient choice to describe the perturbation is the Einstein frame where we have only one metric perturbation
$\Phi_E$ in the absence of anisotropic stress and one scalar perturbation coming from the scalar field. In the conformal Newton gauge, the scalar perturbation can be conveniently represented by the perturbation of the scalar field $\delta\phi$. Other choices such as the Mukhanov-Sasaki variable could have been taken. It turns out the the choice of $\delta\phi$ is more easily implemented numerically. In the numerical applications, we always consider two
Newtonian potentials $\Phi_E$ and $\Psi_E$ as the total anisotropic stress does not always vanish\footnote{The numerical results are obtained in the synchronous gauge and we have taken into account adequately the effect of anisotropic stress in this gauge.}, for instance in the presence of neutrinos.

\subsection{Scalar models}
As already mentioned, we are interested in the effects of scalars mediating a screened modification of gravity on the CMB and Large Scale Structure formation.
We shall work in the Einstein frame\footnote{We consider only models where the scalar field has a mass much larger than the Hubble rate, implying that perturbations in the Jordan and Einstein frames are equivalent.} where the Einstein equations are preserved and depend on the energy momentum tensor of the scalar field :
\begin{equation}
T^\phi_{\mu\nu}= \partial_\mu\phi \partial_\nu \phi -g_{\mu\nu} \left[ \frac{1}{2} (\partial \phi)^2 +V \right].
\end{equation}
We impose that matter couples to both gravity and the scalar field via the metrics
\begin{equation}
g^{(\alpha)}_{\mu\nu} =A^2_{(\alpha)}(\phi) g_{\mu\nu}, \ \ \alpha=b,c
\end{equation}
for each matter species. Their couplings to matter are defined as
\begin{equation}\label{beta_def}
\beta_{\alpha}(\phi)= m_{\rm Pl}\frac{\partial \ln A_{\alpha} (\phi)}{\partial \phi},
\end{equation}
where $m_{\rm Pl}^{-2}\equiv 8\pi G_N$ is the reduced Planck mass, and which may be field dependent in dilatonic models for instance. It is a universal constant $\beta=\frac{1}{\sqrt 6}$ in $f(R)$ models. We will give specific examples later using generalised chameleonic models and models where the coupling function has a transition before recombination.

In all the models that we will consider, the background cosmology is tantamount to a $\Lambda$CDM model since Big Bang Nucleosynthesis (BBN). Deviations from General Relativity only  appear at the perturbative level.
Moreover, the effective potential for the scalar field in the matter era is modified by the presence of matter as a consequence of the non-trivial matter couplings
\begin{equation}
V_{\rm eff}(\phi)= V(\phi) + \sum_{\alpha} \left[A_{\alpha}(\phi)-1\right] \rho_{\alpha},
\end{equation}
where the sum is taken over the non-relativistic species and $\rho_{\alpha}$ is the conserved energy density of the fluid $\alpha$, related to the normal Einstein frame matter density by $\rho_\alpha = \rho^E_\alpha / A_\alpha$. The potential acquires a slowly varying minimum $\phi(\rho_{\alpha})$ as long as the mass $m^2 = \frac{d^2 V_{\rm eff}}{d^2\phi}\vert_{\phi(\rho_{\alpha})}$ is larger than the Hubble rate. We will always assume that this is the case in the following.
Due to the interaction with the scalar field, matter is not conserved and satisfies
\begin{equation}
D_\mu T^{\mu\nu}_{(\alpha)}= \frac{\beta_{\alpha}}{m_{\rm Pl}} \left( \partial^\nu\phi \right) T^{(\alpha)}.
\end{equation}
where $T^{\mu\nu}_{(\alpha)}$ is the energy momentum tensor of the species $\alpha$ and $T^{(\alpha)}$ its trace. Notice that radiation is not affected by the presence of the coupled scalar field.
The Einstein equations are not modified and read
\begin{equation}
R_{\mu\nu}-\frac{1}{2} g_{\mu\nu} R= 8\pi G_N \left(\sum_{\alpha} T^{(\alpha)}_{\mu\nu} +T^\phi_{\mu\nu}\right)~.
\end{equation}
The conservation equations and the Einstein equations are enough to characterise the time evolution of the fields.

\subsection{Tomography}

It is convenient to introduce the total coupling function $A$
\be
A=\sum_\alpha f_\alpha A_\alpha,
\ee
where $f_\alpha=\rho_\alpha/\rho$ is the constant fraction of the species $\alpha$ where $\rho=\sum_{\alpha} \rho_{\alpha}$ is the total conserved energy density of non-relativistic matter. The total coupling $\beta$ is obtained accordingly
\be
\beta A = \sum_{\alpha} f_\alpha \beta_{\alpha}A_{\alpha}.
\ee
Screened modified gravity models satisfy a tomographic description~\cite{Brax:2011aw,Brax:2012gr} whereby the potential $V(\phi)$ and the coupling constants $\beta_{\alpha}(\phi)$ can be reconstructed solely from the knowledge of the density or scale factor dependence of the mass $m(a)$ at the minimum of the effective potential and the total coupling constant $\beta(a)$. At the minimum
\be\label{eq:minimum_condition}
\frac{\mathrm dV}{\mathrm d\phi} = -\frac{\beta A\rho}{m_{\rm Pl}},
\ee
and
\be\label{eq:m_eff}
m^{2}(a)=\frac{\mathrm d^2 V_{\rm eff}}{\mathrm d^2\phi}=\frac{\mathrm d^2 V}{\mathrm d^2\phi}+\frac{\beta^{2}A\rho}{m_{\rm Pl}^{2}}+\frac{A\rho}{m_{\rm Pl}}\frac{\mathrm d\beta}{\mathrm d\phi}.
\ee
Differentiating Eq. \ref{eq:minimum_condition} with respect to conformal time,
\be
\frac{\mathrm d^2 V}{\mathrm d^2\phi}\phi' = -\left(A\rho\frac{\mathrm d\beta}{\mathrm d\phi}+\beta\rho\frac{\mathrm dA}{\mathrm d\phi}\right)\frac{\phi'}{m_{\rm Pl}}-\frac{\beta A\rho'}{m_{\rm Pl}},
\ee
and then using Eq. \ref{eq:m_eff} we find
\be\label{eq:phidot}
\phi' = \frac{3{\cal H}\beta A\rho}{m^{2}(a)m_{\rm Pl}},
\ee
and so, given $\rho =\rho_0 /a^3$,
\be\label{eq:tomography1}
\phi(a)= \phi_c + \frac{3\rho_0}{m_{\rm Pl}}\int_{a_i}^a \frac{\beta (a)}{a^4 m^2(a)} \mathrm da,
\ee
where $\phi_c$ is the field value at the minimum corresponding to the density $\rho(a_i)$ and we have taken $A \approx 1$ (justified by the constraint on the time variation of fermion masses, see section \ref{BBN}). Eq. \ref{eq:minimum_condition} also implies that
\be\label{eq:tomography2}
V(a)= V_0 -\frac{3\rho_0^2}{m_{\rm Pl}^2} \int_{a_i}^a \frac{\beta^2(a)}{a^7 m^2(a)} \mathrm da,
\ee
yielding an implicit definition of $V(\phi)$.

As an example, in the case where $\beta_a=\beta=\frac{1}{\sqrt 6}$, one can reconstruct the large curvature $f(R)$ model (see eq. (\ref{lcm})) and their chameleon mechanism using
\be
m(a)= m_0 \left(\frac{4\Omega_{\Lambda 0}+ \Omega_{m0} a^{-3}}{4\Omega_{\Lambda 0}+ \Omega_{m0}}\right)^{(n+2)/2},
\ee
where the mass on large cosmological scale is given by
\be
m_0= H_0 \sqrt{\frac{4\Omega_{\Lambda 0}+ \Omega_{m0} }{(n+1) f_{R_0}}},
\ee
$\Omega_{\Lambda 0} \approx 0.73$, $\Omega_{m0}\approx 0.27$ are the dark energy and matter density fractions now \cite{Brax:2012gr}.
In the following, we will present other  models defined by the tomographic map.

\subsection{The models}
When the field sits in the minimum of the effective potential, the background follows that closely of a $\Lambda$CDM model, with the evolution of $V(a)\approx {\rm const}$ given in (\ref{eq:tomography2}). But we will see that quite distinctive features can appear when the evolution of perturbations is studied. To be concrete, we will concentrate on two models in this paper, which we will now describe.

\subsubsection{Generalised Chameleons}

In our first  model the scalar field mass and the couplings to baryons and dark matter evolve effectively like power-laws in the scale factor,
\be
m(a) = m_0 a^{-p}, \hspace{1cm} \beta_{\alpha}(a) = \beta_{0\alpha} a^{-b}~.
\ee
Such a model corresponds to generalised chameleons~\cite{Brax:2011aw,Brax:2012gr}. Setting $p \geq 3$, $b=0$ and $\beta_{b0} = \beta_{c0} = 1 / \sqrt 6$, one recovers the large curvature $f(R)$ model \cite{Brax:2011aw,Brax:2012gr}.

We focus here on the case $b > 0$, for which the coupling to matter is very large during the tight coupling regime but smaller at later times. As we shall discuss in section \ref{localtests}, the local tests will impose different constraints on $m_0 $ and $\beta_0$ for different combinations of the exponents $p$ and $b$. One also requires $p \ge 3/2 $ during the matter dominated era, and $p \geq 2$ during the radiation era, so that the scalar field mass is always much larger than the Hubble rate, guaranteeing a $\Lambda$CDM background expansion.

\subsubsection{Transition in $\beta$}

The second model we consider is where the Universe undergoes a smooth but rapid transition from an epoch of strong coupling between the scalar field and matter to one when the coupling is small and has negligible effect on the growth of perturbations \cite{vandeBruck:2012vq,Brax:2013fna}. Using the tomographic maps allowing one to reconstruct $V(\phi)$ and $\beta(\phi)$, one finds that the potential is an inverse power law both before and after a transition point $\phi_{\rm trans}$ where the potential decreases abruptly. The effective potential, both before and after the transition, has a minimum with a large mass where the field gets trapped. Dynamically, the field undergoing the transition jumps from the minimum before the transition to the minimum after the transition where it will oscillate a few times before being rapidly (the amplitude decreases like $a^{-3/2}$) stuck at the minimum again. We will neglect these decaying oscillations in the following and assume that the field tracks the minimum at all times (for a similar type of phenomenon, see the thorough analysis of the transition in \cite{Brax:2011pk}. In order to satisfy the constraints imposed by local tests we are considering very high masses and this will ensure that the field remains close to the minimum through the transition.

We can parameterise this behaviour with the function
\be
\beta_{c,b}(a) = \beta_0 + \frac{\beta_{i}}{2}\left[1 + \rr{tanh} ( C ( \atrans - a ) ) \right]~,
\ee
where the parameter $C$ controls the duration of the transition.
We choose the effective mass of the scalar field to evolve like a power-law
\be
m(a) = m_0 a^{-p}.
\ee
By setting the transition prior to last scattering, we ensure that all the effects on the CMB angular power spectrum and on the matter power spectrum are mainly caused by the modification of gravity before recombination. Note however that the linear perturbations at recombination, which can be used as initial conditions for the growth of the matter perturbations, are modified and thus can lead to a different evolution compared to the $\Lambda$CDM model, even if there is no direct effect of modified gravity on the growth of structures. For the parameters we have considered, values of $\beta_0 \lesssim 10^{-4} m_0/H_0$ do not lead to any visible modification in the growth of perturbations after last scattering. Since we are considering rather high masses this means a strong coupling even up to the present epoch is not ruled out.

\subsection{BBN Constraint}\label{BBN}

In all the models that we consider the masses of fundamental particles vary as
\be
m_\psi = A(\phi)m_{\rm bare},
\ee
where $m_{\rm bare}$ is the bare mass appearing in the matter Lagrangian. The measurements of primordial light element abundances place a tight constraint on the time variation of fermion masses since BBN~\cite{Olive:2007aj,Bedaque:2010hr,Berengut:2009js,Nesseris:2009jf}
\be
\frac{\Delta m_\psi}{m_\psi} = \frac{\Delta A}{A} \lesssim 0.1.
\ee
Therefore we must require that $A \approx 1$ since BBN. Using \eqref{beta_def} and \eqref{eq:phidot} we get
\be
\frac{dA}{da} = \frac{3\beta^{2}\rho}{am^{2}m_{\rm Pl}^{2}},
\ee
and so
\be
\Delta A = 9\Omega_{m,0}H_0^2 \int_{a_{\rm BBN}}^{a_0}\frac{\beta^{2}(a)}{a^{4}m^{2}(a)}da \lesssim 0.1.
\ee
For chameleon models this is
\be
\frac{9\Omega_{m,0}H_0^2 \beta_0^2}{m_0^2}\frac{(a_0^{2p-2b-3}-a_{\rm BBN}^{2p-2b-3})}{2p-2b-3} \lesssim 0.1.
\ee
For the case $2p-2b-3>0$ this yields
\be
\frac{\beta_0}{m_0} \lesssim \left(\frac{2p-2b-3}{90\Omega_{m,0}H_0^2}\right)^{1/2} \lesssim \mathcal O(10^3)\Mpc,
\ee
while for $2p-2b-3=0$
\be
\frac{\beta_0}{m_0} \lesssim (90\Omega_{m,0}H_0^2 {\rm ln}10^{9})^{-1/2} \lesssim \mathcal O(10^2)\Mpc.
\ee
However if $2p-2b-3<0$ we find
\be
\frac{\beta_0}{m_0} \lesssim \left(\frac{|2p-2b-3|}{9\Omega_{m,0}H_0^2}\right)^{1/2}10^{\frac{9(2p-2b-3)-1}{2}} \lesssim \mathcal O(10^{\frac{6+9(2p-2b-3)}{2}})\Mpc,
\ee
which results in a much tighter constraint on $\frac{\beta_0}{m_0}$ than that coming from the local tests.

For the model with a transition in $\beta$ the constraint is
\be
\frac{9\Omega_{m,0}H_0^2}{m_0^2 (2p-3)}[\beta_{\rm ini}^2 (a_{\rm trans}^{2p-3}-a_{BBN}^{2p-3})+\beta_0^2 (a_0^{2p-3}-a_{\rm trans}^{2p-3})] \lesssim 0.1.
\ee
We shall only consider cases where $2p-3>0$ so that
\be
\frac{9\Omega_{m,0}H_0^2 \beta_{\rm ini}^2}{m_0^2 (2p-3)}a_{\rm trans}^{2p-3} \lesssim 0.1,
\ee
and if we let $\atrans=10^{-t}$ we find
\be
\frac{\beta_{\rm ini}}{m_0} \lesssim \left(\frac{2p-3}{9\Omega_{m,0}H_0^2}\right)^{-1/2}10^{\frac{t(2p-3)-1}{2}} \lesssim \mathcal O(10^{3+t(p-\tfrac{3}{2})})\Mpc.
\ee
Typically, we will see that this constraint is superseded by the local constraints.

\subsection{Local tests}\label{localtests}

We will now examine the constraints on our two models coming from local tests of gravity. We focus on three tests which can yield strong constraints on screened models. The first comes from the requirement that the Milky Way should be screened in order to avoid large, disruptive effects on its dynamics~\cite{Pourhasan:2011sm,Jain:2012tn}. The second is the laboratory tests of gravity involving spherical bodies which should not feel a large fifth force~\cite{Adelberger:2003zx,Adelberger:2002ic}. The last one follows from the lunar ranging experiment~\cite{Williams:2012nc,Williams:2004qba,Williams:2003wu,Williams:2005rv} which looks for violations of the strong equivalence principle, and which turns out to be the strongest constraint for the models we have chosen.

The screening condition is a simple algebraic relation
\be
\vert \phi_{\rm in} -\phi_{\rm out}\vert \le 2\beta_{\rm out} m_{\rm Pl} \Phi_N~,
\ee
where $\Phi_N$ is Newton's potential at the surface of a body and $\phi_{\rm{in,out}}$ are the values of the field inside and outside the body.
Using the tomography equation (\ref{eq:tomography1}), we find the relation
\begin{equation}
\frac{\vert \phi_{\rm in} -\phi_{\rm out}\vert}{m_{\rm Pl}}=  \frac{3}{m_{\rm Pl}^2}\int_{a_{\rm  in}}^{a_{\rm out}} \frac{\beta(a) }{a m^2(a)}\rho (a)  da ,\label{phi}
\end{equation}
where $a_{\rm in,out}$ are defined by
$
\rho (a_{\rm in, out})= \rho_{\rm in,out}~$
This condition expresses the fact that the effective modification of Newton's constant felt by an unscreened body in the presence of a screened object due to the scalar field goes from
$2\beta^2$ to $2 \beta^2 \frac{\vert \phi_{\rm in} -\phi_{\rm out}\vert}{ 2\beta m_{\rm Pl} \Phi_N}$.
When two screened bodies such as the moon and the earth fall in the gravitational field of the sun, the scalar field induces a relative acceleration between them which is measured by the square of their scalar charges
\be
Q=\frac{\vert \phi_{\rm in} -\phi_{\rm out}\vert}{ m_{\rm Pl} \Phi_N}
\ee
and the lunar ranging experiment requires that the earth's charge $Q_\oplus \le 10^{-7}$ where $\phi_\oplus\sim 10^{-9}$.

For the Milky Way, the density inside the galaxy is typically $10^6$ times the cosmological matter density now, i.e. $a_G\sim 10^{-2}$. The value outside the galaxy is the cosmological one (if it is in a dense cluster then $a_{\rm out} <1$ and the screening condition is less stringent) so the screening condition reads
\begin{equation}
\frac{\vert \phi_{\rm G} -\phi_{0}\vert}{m_{\rm Pl}}=  \frac{3}{m_{\rm Pl}^2}\int_{a_{G}}^{a_0} \frac{\beta(a) }{a m^2(a)}\rho (a)  da \le 2\beta_{G} \Phi_G
\end{equation}
where $\Phi_G\sim 10^{-6}$. For the generalised chameleon model, this is
\be
\frac{9 \Omega_{m0} \beta_0 H_0^2}{(2p-b-3)m_0^2}(a_0^{2p-b-3}-a_G^{2p-b-3}) \le 2 \beta_{G} \Phi_G~.
\ee
When $2p-b-3<0$, the contribution from $a_G$ dominates and one gets
\be
\frac{m_0^2}{H_0^2} \gtrsim \frac{9 \Omega_{m0}  }{2 |2p+b-3| \Phi_G} a_G^{2p-3} \gtrsim \mathcal O(10^{6-2(2p-3)})~.
\ee
If on the other hand $2p-b-3>0$, the contribution from $a_0$ dominates and, setting $a_0 \equiv 1$ as usual, one has
\be
\frac{m_0^2}{H_0^2} \gtrsim \frac{9 \Omega_{m0}  }{2 (2p-b-3) \Phi_G} \gtrsim \mathcal O(10^{6})~.
\ee
For the transition model with $\atrans<a_G$, one gets a similar constraint,
\be
\frac{m_0^2}{H_0^2} \gtrsim \frac{9 \Omega_{m0}  }{2 (2p-3) \Phi_G} \gtrsim \mathcal O(10^{6})~.
\ee
which is independent of the value of $\beta_0$.

Tests of gravity in cavities involve spherical bodies of order 10~cm, with a Newtonian potential of order $\Phi_c\sim 10^{-27}$ and a density $\rho_{c}\sim 10~{\rm g}$ \cite{Khoury:2003rn}. This density is of order of the cosmological density just before BBN when the scalar field must have settled at the minimum of the effective potential. Outside, in the cavity, the scalar field takes a value such that its mass is of order $m_{\rm cav} L \sim 1$ corresponding to a scale factor  $a_{\rm cav}= (L m_0)^{1/p}$. The constraint reads
\begin{equation}
\frac{\vert \phi_{\rm c} -\phi_{\rm cav}\vert}{m_{\rm Pl}}=  \frac{3}{m_{\rm Pl}^2}\int_{a_{c}}^{a_{\rm cav}} \frac{\beta (a)}{a m^2(a)}\rho (a)  da \le 2\beta_c \Phi_c ~,
\end{equation}
which for the chameleon model with $2p-b-3>0$ is
\be
\frac{m_0^2}{H_0^2} \gtrsim \frac{9\Omega_{m0}}{2\Phi_c}\frac{a_{\rm cav}^{2p-b-3}}{2p-b-3}a_{c}^{b} \gtrsim \mathcal O(10^{2})~.
\ee
For the transition model we find
\be
\frac{m_0^2}{H_0^2} \gtrsim \frac{9\Omega_{m0}}{2\Phi_c}\frac{a_{\rm cav}^{2p-3}}{2p-3} \gtrsim \mathcal O(10^{8})~,
\ee
which is a tighter constraint than that coming from the screening of the Milky Way.

A much more stringent condition follows from the lunar ranging experiment,
\begin{equation}
\frac{3}{m_{\rm Pl}^2}\int_{a_c}^{a_G} \frac{\beta(a) }{a m^2(a)}\rho (a)  da \le Q_\oplus \Phi_\oplus~,
\end{equation}
where $ Q_\oplus \Phi_\oplus\sim 10^{-16}$ and $a_c$ is associated with the densities inside the earth. For the generalised chameleon model the case $2p - b - 3 < 0 $ leads to
\be
\frac{m_0^2}{H_0^2} \gtrsim \frac{9 \Omega_{m0} \beta_0 }{ |2p-b-3|  Q_\oplus \Phi_\oplus} a_c^{2p-b-3}~.
\ee
Since $a_c\ll 1$, considering values for which $ 2p - b - 3 \lesssim -1 $ implies an extremely stringent constraint on the ratio $m_0^2 /\beta_0$ that precludes any visible cosmological signature of the model. We therefore focus on the opposite case  $2p - b - 3 > 0 $ for which one gets
\be
\frac{m_0^2}{H_0^2} \gtrsim \frac{9 \Omega_{m0} \beta_0 }{Q_\oplus \Phi_\oplus} \frac{a_G^{2p-b-3}}{ 2p-b-3}
\gtrsim \beta_0 \mathcal O(10^{16-2(2p-b-3)})~.
\ee
For cases where $\beta_0 \sim \mathcal O(1)$, $ b \simeq0$ and $p \gtrsim 3$, the lunar ranging bound on $m_0$ is
weaker than the constraint from the Milky Way. For the transition model, assuming $\atrans<a_G$ and $2p-3>0$, the condition is
\be
\frac{m_0^2}{H_0^2} \gtrsim \frac{9 \Omega_{m0} \beta_i }{Q_\oplus \Phi_\oplus} \frac{a^{2p-3}_{\rr{trans}}}{2p-3} \gtrsim \beta_i a^{2p-3}_{\rr{trans}} \mathcal O(10^{16})~.
\ee
For the case $p=3$, $\beta_{i}=10^{13}$ and $\ztrans=1090$, this gives the constraint $m_0 \gtrsim 10^{10}H_0 $.

\section{Perturbations}

In this section, we study the evolution of perturbations with analytical methods, to gain an understanding of the physics involved. From now on we will work in natural units where $8\pi G_N = 1$. We are interested in the first order perturbations of the Einstein equation and the conservation equations. In the absence of anisotropic stress, the metric can be described in the conformal Newton gauge
\begin{equation}
\mathrm ds^2= a^2(\eta) \left[-(1+2\Phi) \mathrm d\eta^2 + (1-2\Phi) \mathrm dx^2 \right],
\end{equation}
where $\Phi$ is Newton's potential. In this gauge the relevant Einstein equations are (in Fourier space)
\begin{equation} \label{eq:Phi}
k^2\Phi + 3 {\cal H}( \Phi' + {\cal H} \Phi)= -a^2 \sum_{\alpha} \delta\rho_{\alpha} -a^2 \left(\frac{\phi'^2}{a^2}\Phi - \frac{\phi'}{a^2}\delta\phi' - V_{,\phi}\delta\phi\right)
\end{equation}
and
\begin{equation}
k^2 (\Phi' + {\cal H} \Phi)= a^2 \sum_{\alpha} (1+ w_{\alpha}) \theta_{\alpha} + k^2 \phi' \delta \phi~,
\end{equation}
where $\delta\rho_{\alpha} =\rho^E_{\alpha}\delta_{\alpha}$ is the Einstein frame density perturbation. The Hubble rate is ${\cal H}= a'/a$ and the equation of state of each species is $w_{\alpha}$. We have defined the divergence of the velocity field of each species $\theta= \partial_i v^i$.
The perturbed Klein-Gordon equation reads
\begin{equation} \label{eq:pertKG}
\delta\phi'' +2{\cal H} \delta \phi' +(k^2 + a^2 V_{,\phi\phi})\delta \phi + 2\Phi a^2  V_{\rm eff,\phi}-4\Phi' \phi'+ a^2 \left(\beta_c \delta\rho_c +\beta_b \delta\rho_b \right)+ a^2 \left(\beta_{c,\phi}\rho^E_c +\beta_{b,\phi}\rho^E_b \right)\delta\phi=0~.
\end{equation}

We now consider the perturbation equations in CDM, baryons and radiation.
The conservation equation for CDM leads to the coupled equations
\begin{equation} \label{eq:deltac}
\delta_c' = -(\theta_c -3 \Phi') + \beta_{c,\phi}\phi' \delta\phi + \beta_c \delta\phi',
\end{equation}
and
\begin{equation}  \label{eq:thetac}
 \theta_c'= -{\cal H} \theta_c + k^2 \Phi + k^2 \beta_c \delta \phi -\beta_c \phi' \theta_c.
\end{equation}
For the baryons we get
\begin{equation} \label{eq:deltab}
\delta_b'=-\theta_b +3 \Phi' + \beta_{b,\phi}\phi' \delta\phi + \beta_b \delta\phi',
\end{equation}
and
\begin{equation}  \label{eq:thetab}
\theta_b'= -{\cal H}\theta_b +\frac{an_e \sigma_T}{R} (\theta_\gamma-\theta_b)+ k^2\Phi + \beta_b k^2 \delta \phi -\beta_b \phi' \theta_b,
\end{equation}
where we have added the interaction term to account for the coupling to photons via Thompson scattering and $R=3\rho^E_b/4\rho_\gamma$.

The Thompson scattering cross section depends on the electron mass $m_e$ which is field dependent due to the conformal rescaling of the metric, i.e. $m_e$ is proportional to $A_b(\phi)$. However, since the scalar field tracks the minimum of the effective potential, the time variation of the electron mass in one Hubble time is suppressed by ${\cal O} (\frac{H^2}{m^2}) \ll 1$ and can therefore be neglected.

To describe the photons we shall work in the fluid approximation and since the Boltzmann hierarchy is not altered by the presence of a scalar field we have
\begin{equation} \label{eq:deltagamma}
\delta_\gamma'= -\frac{4}{3} \theta_\gamma +4 \Phi'
\end{equation}
and
\begin{equation} \label{eq:thetagamma}
\theta_\gamma'= \frac{k^2}{4} \delta_\gamma + k^2 \Phi + an_e \sigma_T (\theta_b-\theta_\gamma).
\end{equation}

We also need to specify the initial conditions for all the perturbations. We will be interested in modes which will enter the horizon before radiation-matter equality. Adiabatic initial conditions are determined by
$\delta_c^i= \delta_b^i= \frac{3}{4} \delta_\gamma^i $ and $\delta_b^i= -\frac{3}{2} \Phi^i $.

On subhorizon scales and neglecting the time variation of $\phi$, the equations simplify:
\begin{equation} \label{eq:deltacsubH}
\delta_c' = -\theta_c ,
\end{equation}
\begin{equation}  \label{eq:thetacsubH}
\theta_c'= -{\cal H} \theta_c + k^2 \Phi + k^2 \beta_c \delta \phi.
\end{equation}
\begin{equation} \label{eq:deltabsubH}
\delta_b'=-\theta_b ,
\end{equation}
\begin{equation}  \label{eq:thetabsubH}
\theta_b'= -{\cal H}\theta_b +\frac{an_e \sigma_T}{R} (\theta_\gamma-\theta_b)+ k^2\Phi + \beta_b k^2 \delta \phi,
\end{equation}
\begin{equation} \label{eq:deltagammasubH}
\delta_\gamma'= -\frac{4}{3} \theta_\gamma
\end{equation}
\begin{equation} \label{eq:thetagammasubH}
\theta_\gamma'= \frac{k^2}{4} \delta_\gamma + k^2 \Phi + an_e \sigma_T (\theta_b-\theta_\gamma).
\end{equation}

To leading order in the tight-coupling approximation $an_{e}\sigma_{T}\to\infty$ which implies $\theta_b\approx \theta_\gamma$, and therefore the photon and baryon density contrasts are linked by $\delta_b\approx \frac{3}{4}\delta_\gamma$. This leads to
\be
\delta_b'' = -\frac{R'}{(1+R)}\delta_b'-k^{2}c_s^{2}\delta_b-k^{2}\Phi-\frac{R}{(1+R)}\beta_b k^2\delta\phi,
\ee
where $c_{s}=1/\sqrt{3(1+R)}$ is the standard sound-speed. The Klein-Gordon equation in the subhorizon limit is
\be
\delta \phi = -\frac{\beta_c \delta\rho_c +\beta_b \delta\rho_b } {\frac{k^2}{a^2}+V_{,\phi\phi}+\beta_{c,\phi}\rho^E_c +\beta_{b,\phi}\rho^E_b }. \label{eq:KGsubH}
\ee
Using this and approximating the Poisson equation with $2k^2\Phi=-a^2\delta\rho_c$, and then defining $\delta_b =(1+R)^{-1/2} \delta$ yields
\be
\delta'' + c_s^2 k^2 \left(1- \frac{9\Omega_b \beta_b^2 R{\cal H}^2 }{{k^2} +a^2V_{,\phi\phi}+3{\cal H}^2(\beta_{c,\phi}\Omega_{c}+\beta_{b,\phi}\Omega_{b})}\right)\delta=-k^2(1+R)^{1/2}\left(1+ \frac{2\beta_b \beta_c}{1+\frac{a^2V_{,\phi\phi} +3{\cal H}^2(\beta_{c,\phi}\Omega_{c}+\beta_{b,\phi}\Omega_{b})}{k^2}} \frac{R}{R+1}\right)\Phi.
\ee
This can be simplified by introducing
\begin{equation} \label{deltatilde}
\tilde \delta= \delta + (1+R)^{1/2}\left(1+ \frac{2\beta_b \beta_c}{1+\frac{a^2V_{,\phi\phi} +3{\cal H}^2(\beta_{c,\phi}\Omega_{c}+\beta_{b,\phi}\Omega_{b})}{k^2}} \frac{R}{R+1}\right)\frac{\Phi}{\tilde c_s^2},
\end{equation}
where the effective speed of sound is
\begin{equation}\label{eq:effsoundspeed}
\tilde c_s^2= c_s^2 \left(1- \frac{9\Omega_b \beta_b^2 R{\cal H}^2 }{{k^2} +a^2V_{,\phi\phi}+3{\cal H}^2(\beta_{c,\phi}\Omega_{c}+\beta_{b,\phi}\Omega_{b})}\right).
\end{equation}
This leads to
\begin{equation}
\tilde \delta'' +\tilde k^2 c_s^2 \tilde\delta=0,
\end{equation}
and using the WKB method the solution is
\begin{equation}
\tilde \delta= \tilde c_s^{-1/2}B\cos k\tilde r_s,
\end{equation}
where $B$ is set by the initial conditions and $\tilde r_s(\eta)= \int_0^\eta \tilde c_s d\eta$ is the modified sound horizon. We can relate this to back to the baryon perturbation,
\be\label{eq:deltabsoln}
\delta_b = \frac{\tilde \delta}{(1+R)^{1/2}} - \left(1+ \frac{2\beta_b \beta_c}{1+\frac{a^2V_{,\phi\phi} +3{\cal H}^2(\beta_{c,\phi}\Omega_{c}+\beta_{b,\phi}\Omega_{b})}{k^2}} \frac{R}{R+1}\right)\frac{\Phi}{\tilde c_s^2}.
\ee
We should note that the WKB approximation is only valid if the time variation of $\tilde c_s$ is smooth. This will be violated for a short period in the models with a transition in $\beta$.

As implied above the evolution of the Newtonian potential before last scattering is dominated by that of the CDM perturbations. Equations \eqref{eq:deltacsubH} and \eqref{eq:thetacsubH} therefore lead to the growth equation for CDM
\begin{equation}\label{eq:deltac''}
\delta_c'' +{\cal H} \delta_c' -\frac{3}{2} {\cal H}^2 \Omega_c \left(1+ \frac{2\beta_c^2}{1+\frac{a^2V_{,\phi\phi} +3{\cal H}^2(\beta_{c,\phi}\Omega_{c}+\beta_{b,\phi}\Omega_{b})}{k^2}}\right)\delta_c-\frac{3{\cal H}^2 \Omega_b\beta_b\beta_c}{1+\frac{a^2V_{,\phi\phi} +3{\cal H}^2(\beta_{c,\phi}\Omega_{c}+\beta_{b,\phi}\Omega_{b})}{k^2}}\,\delta_b=0
\end{equation}
As noted in \cite{BraxDavisMGCMB} we can identify three possible sources of modification to the CMB angular power spectrum. These are the modified sound horizon which could cause a shift in the peak positions, the modified evolution of the Newtonian potential due to anomalous growth of the CDM perturbations, and an extra contribution to the growth of the baryon perturbations proportional to $\Phi$ and the couplings to both baryons and CDM.

\section{Numerical results: CMB angular power spectrum and Matter power spectrum}

We have implemented the linear perturbation dynamics for two models of screened gravity within a modified version of the CAMB code \cite{CAMB}. The details of the modifications that were required are given in the appendix.

\subsection{Transition in $\beta$}

\begin{figure}
\begin{center}
\includegraphics[width=13cm]{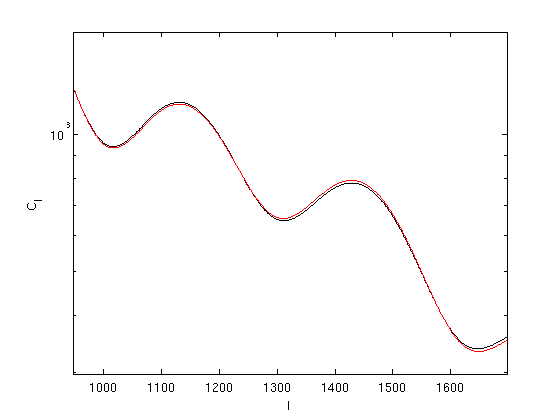}
\end{center}
\caption{CMB angular power spectra of the transition model with $\ztrans = 1090$ and $\beta_i/m_0 = 5\times 10^{6} \Mpc$ (red) and a $\Lambda$CDM model (black), zooming in on the 4th and 5th peaks.}\label{fig:Clzoom2z1090}
\end{figure}

\begin{figure}
\begin{center}
\includegraphics[width=13cm]{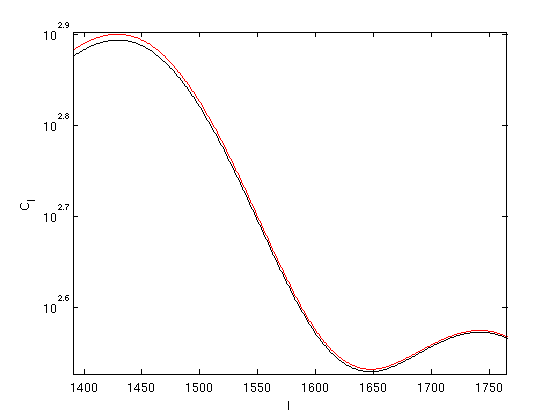}
\end{center}
\caption{CMB angular power spectra of the transition model with $\ztrans = 3000$ and $\beta_i/m_0 = 5\times 10^{7} \Mpc$ (red) and a $\Lambda$CDM model (black), zooming in on the 5th and 6th peaks.}\label{fig:Clzoom2z3000}
\end{figure}

For the transition model we observe deviations from $\Lambda$CDM in both the CMB angular power spectrum and the matter power spectrum which increase with increasing $\beta_i/m_0$. Keeping the ratios $\beta_i / m_0$ and $\beta_0 / m_0$ constant (with the other parameters fixed) results in identical effects. In the CMB the nature of the deviations varies with $l$ and there are alternating periods of enhancement and reduction of power. We see no shift in the positions of the peaks (see Figs. \ref{fig:Clzoom2z1090} and \ref{fig:Clzoom2z3000}). This is because the sound horizon is virtually unchanged. Consequently there is also no shift in the position of the baryon acoustic peaks in the matter power spectrum.

\begin{figure}
\begin{center}
\includegraphics[width=8cm]{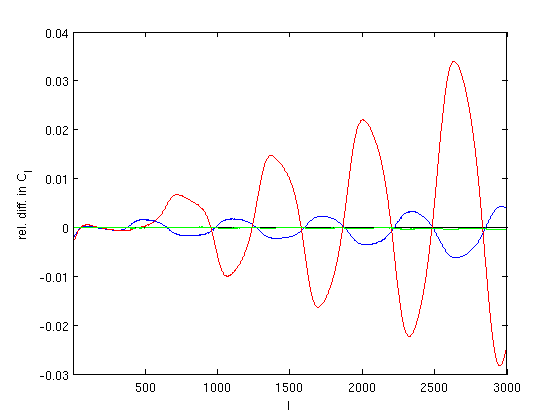}~~~~\includegraphics[width=8cm]{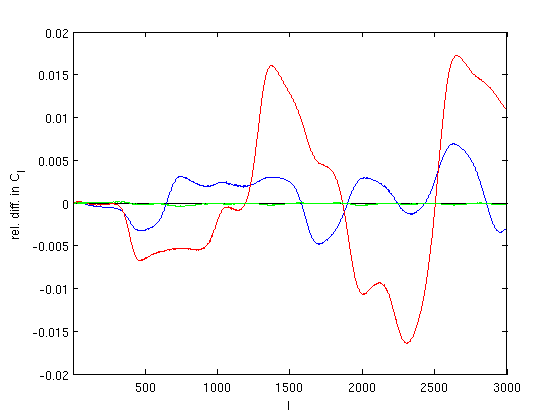}
\includegraphics[width=8cm]{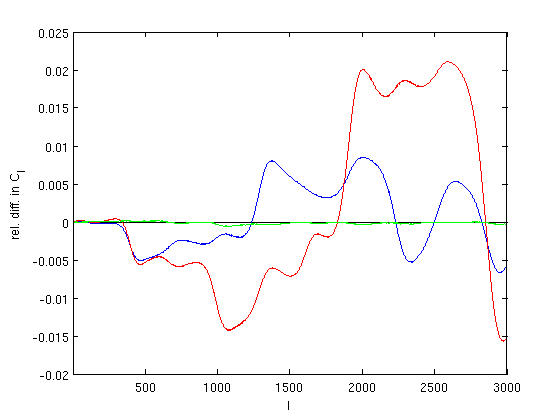}~~~~\includegraphics[width=8cm]{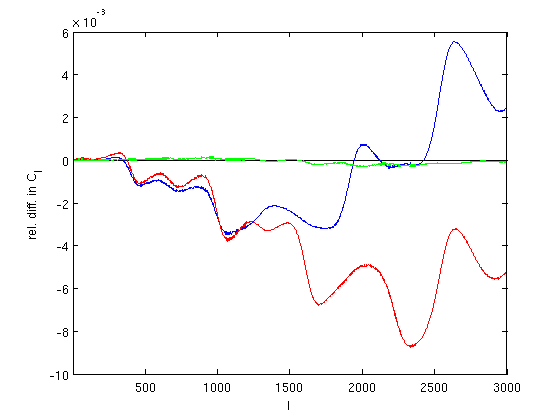}
\end{center}
\caption{Relative differences between the $C_l$'s of the transition models and that of a $\Lambda$CDM model. Red: $\beta_b = \beta_c$, blue: $\beta_b = 0$, green: $\beta_c = 0$. Top left: $\ztrans = 1090$ and $\beta_i/m_0 = 5\times 10^{6} \Mpc$; top right: $\ztrans = 3000$ and $\beta_i/m_0 = 5\times 10^{7} \Mpc$; bottom left: $\ztrans = 5000$ and $\beta_i/m_0 = 1.67\times 10^{8} \Mpc$; bottom right: $\ztrans = 10^4$ and $\beta_i/m_0 = 5\times 10^{8} \Mpc$. For all curves $\beta_0 = 1$, $p=3$ and $C = 10\ztrans$.}\label{fig:Clrdffs}
\end{figure}

The exact effects of the transition model on the angular power spectrum are different depending on when the transition occurs. In Fig. \ref{fig:Clrdffs} we illustrate this with 4 different transition redshifts. For a transition at last scattering ($\ztrans = 1090$) we see clear oscillations in the relative difference of the $C_l$'s whose amplitude increases with increasing $l$. This corresponds to enhanced amplitudes of the odd peaks and troughs and reduced amplitudes of the even peaks and troughs as can be seen in Fig. \ref{fig:Clzoom2z1090}. For $\beta_i/m_0 = 5 \times 10^{6} \Mpc$ the deviations exceed the percent level. In the cases $\ztrans = 3000$ and $\ztrans = 5000$ we see apparently periodic alternation of enhancement and reduction of power. Increasing the transition redshift lengthens these periods. This can be seen in Fig. \ref{fig:Clzoom2z3000} which shows that for $\ztrans = 3000$ two consecutive peaks are enhanced. On average the deviations are greater on smaller scales and the enhancements larger than the reductions. There are also oscillatory features within the intervals of increased and decreased power. Values of $\beta_i/m_0 = 5\times 10^{7} \Mpc$ for $\ztrans = 3000$ and $\beta_i/m_0 = 1.67\times 10^{8} \Mpc$ for $\ztrans = 5000$ produce deviations of around $1\%$. Finally, for $\ztrans = 10000$ there is only a slight oscillating enhancement of the relative difference for $l \lesssim 400$ (just before the first trough) and then a more significant reduction in power for $l \gtrsim 400$ with irregular oscillations in the relative difference. At this redshift the deviations approach the percent level for $\beta_i/m_0 = 5 \times 10^{8} \Mpc$.

Fig. \ref{fig:Clrdffs} also displays the relative importance of the couplings to CDM and baryons for the effects on the CMB. If we set $\beta_b = 0$ the deviations are smaller (generally no more than half a percent) but visible. The oscillations in the relative difference are out of phase with those of the $\beta_b = \beta_c$ case but there is a greater similarity between the effects for different transition redshifts than with $\beta_b = \beta_c$. On the other hand if we set $\beta_c = 0$ the deviations are completely negligible in all cases. These results suggest that the dominant contribution to the modified effects is the term in Eq. \eqref{eq:deltabsoln} which contains both $\beta_c$ and $\beta_b$ and causes the modifications to the growth of the baryon perturbations before last scattering.

\begin{figure}
\begin{center}
\includegraphics[width=8cm]{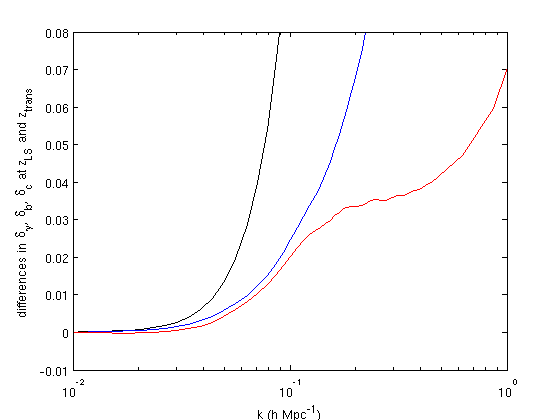}~~~~\includegraphics[width=8cm]{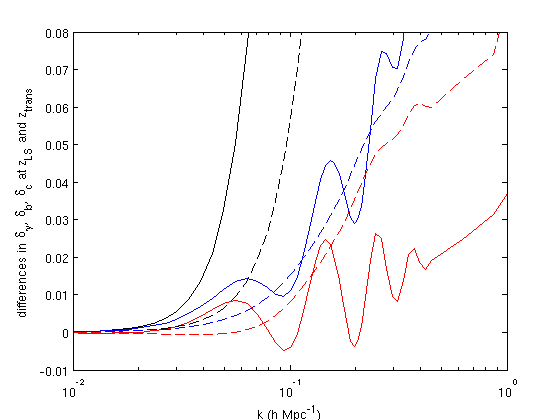}
\includegraphics[width=8cm]{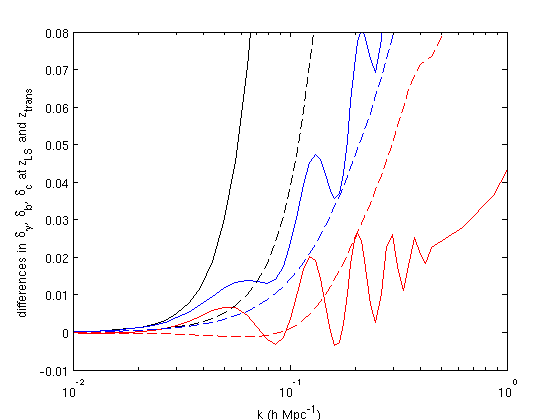}~~~~\includegraphics[width=8cm]{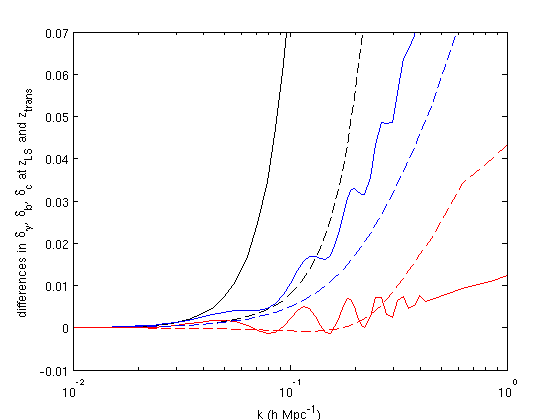}
\end{center}
\caption{Differences between the photon (red), baryon (black) and CDM (blue) transfer functions at $\ztrans$ (dashed) and $z_{LS}=1090$ (solid) of the transition models and a $\Lambda$CDM model. Top left: $\ztrans = 1090$ and $\beta_i/m_0 = 5\times 10^{6} \Mpc$, top right: $\ztrans = 3000$ and $\beta_i/m_0 = 5\times 10^{7} \Mpc$, bottom left: $\ztrans = 5000$ and $\beta_i/m_0 = 1.67\times 10^{8} \Mpc$, bottom right: $\ztrans = 10^4$ and $\beta_i/m_0 = 5\times 10^{8} \Mpc$. For all curves $\beta_0 = 1, p=3$ and $C = 10\ztrans$.}\label{fig:transfers}
\end{figure}

\begin{figure}
\begin{center}
\includegraphics[width=7.5cm]{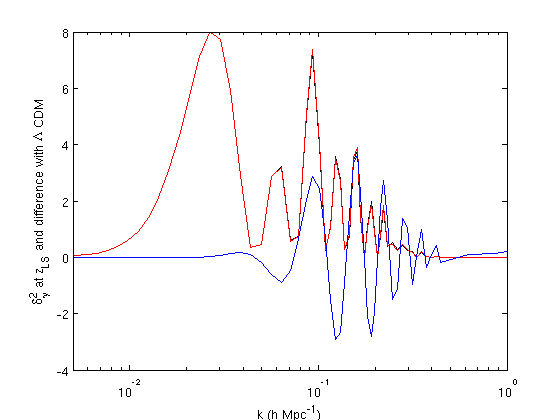}
\includegraphics[width=7.5cm]{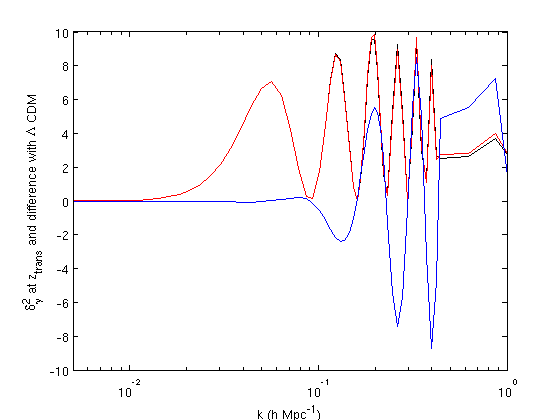}~~~~\includegraphics[width=7.5cm]{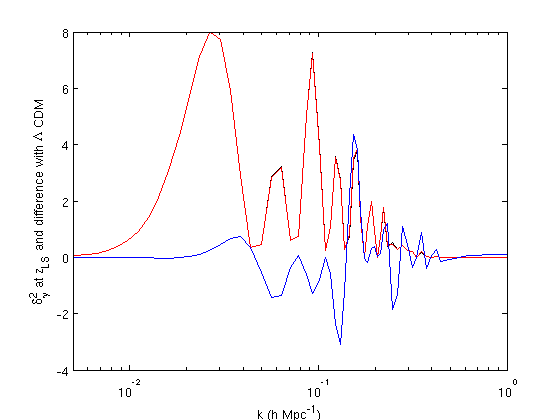}
\includegraphics[width=7.5cm]{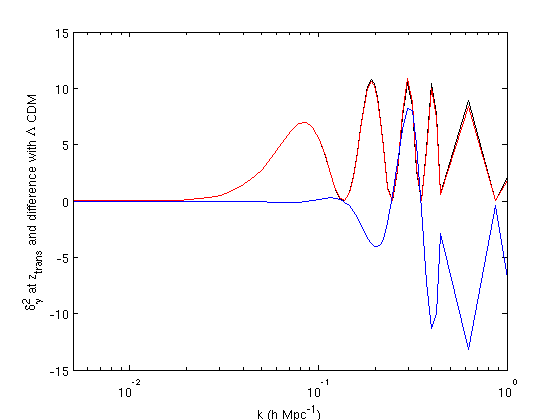}~~~~\includegraphics[width=7.5cm]{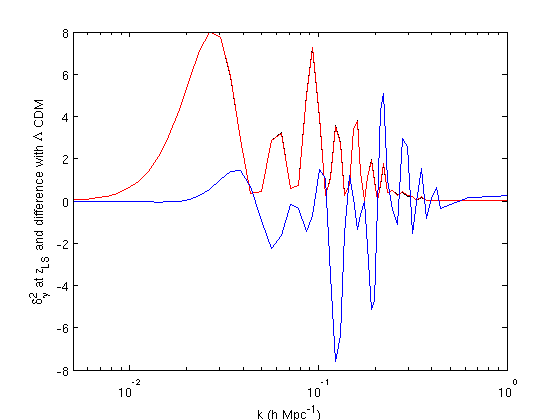}
\includegraphics[width=7.5cm]{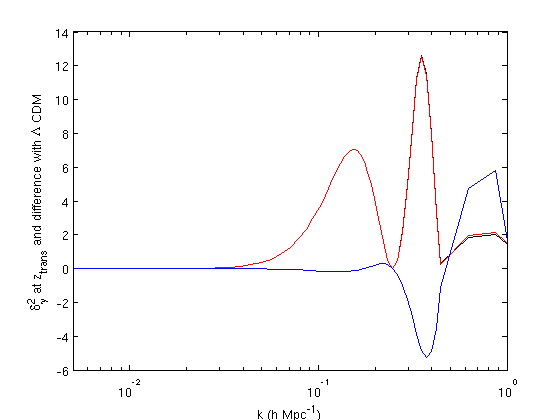}~~~~\includegraphics[width=7.5cm]{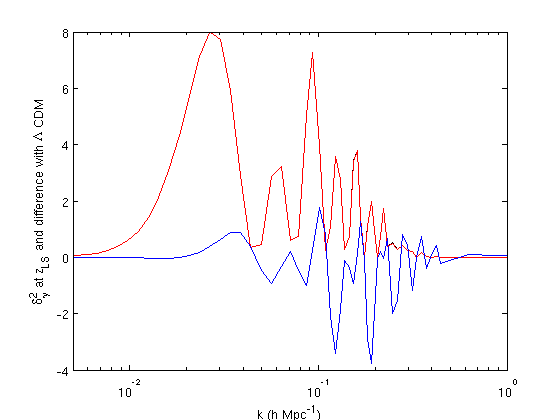}
\end{center}
\caption{$\delta_{\gamma}^2$ for the transition models (red) and a $\Lambda$CDM model (black) at $\ztrans$ (left column) and $z_{LS}$ (right column) and the magnified difference with $\Lambda$CDM (blue). Top: $\ztrans = 1090$ and $\beta_i/m_0 = 5\times 10^{6} \Mpc$, second row: $\ztrans = 3000$ and $\beta_i/m_0 = 5\times 10^{7} \Mpc$, third row: $\ztrans = 5000$ and $\beta_i/m_0 = 1.67\times 10^{8} \Mpc$, bottom row: $\ztrans = 10^4$ and $\beta_i/m_0 = 5\times 10^{8} \Mpc$. For all curves $\beta_0 = 1, p=3$ and $C = 10\ztrans$.}\label{fig:deltagsq}
\end{figure}

The simplest case to understand is that of the transition at last scattering since the evolution of perturbations is governed by the same modified equations right up to the creation of the CMB. In Fig. \ref{fig:transfers} we show the effect of modified gravity on the transfer functions. Superimposed on the oscillations in the photon transfer function that lead to the CMB anisotropies is an enhancement of power which increases with $k$. The peaks and troughs in the angular power spectrum correspond to those in $\delta_{\gamma}^2$ (shown in Fig. \ref{fig:deltagsq}). We see that the odd peaks, corresponding to maxima in $\delta_\gamma$, are enhanced with respect to $\Lambda$CDM whilst the even ones, corresponding to minima in $\delta_\gamma$, are reduced. So far this agrees with what we observe in the $C_l$'s for the $\ztrans = 1090$ case. The troughs in $\delta_{\gamma}^2$ come from the zeros in $\delta_{\gamma}$ and so at these points the difference with $\Lambda$CDM vanishes. However, as we observed above the troughs in the $C_l$'s are alternately enhanced and reduced just like the peaks. We can understand this by noting that the $C_l$ troughs, unlike those in $\delta_{\gamma}^2$, are not zeros. This is because the anisotropy at a given $l$ is created by many modes with wavenumbers greater than that of the principal corresponding $k$-mode. The biggest contribution though will come from modes with only slightly larger $k$ which explains why the troughs preceding odd peaks are also higher and those preceding even peaks are also lower.

When the transition occurs before last scattering the picture is more complicated. The periods of enhancement and reduction of power in the $C_l$'s are different. As can be seen in Fig. \ref{fig:transfers} the effect of modified gravity on the photon transfer function at last scattering is no longer positive on all scales. There are instead oscillations in the difference with $\Lambda$CDM and these are out of phase with those in the transfer function itself. This results in a more complicated and unpredictable pattern of enhancements and reductions of power on different scales in the $\delta_{\gamma}^2$ spectrum (see Fig. \ref{fig:deltagsq}, right) and therefore the CMB. If however we look at the photon transfer functions at $\ztrans$ (Fig. \ref{fig:deltagsq}, left) we see that they are always enhanced compared to the $\Lambda$CDM case and the effects on $\delta_{\gamma}^2$ are the same as those for the transition at last scattering. The only exception to this is the case with $\ztrans = 5000$ where, because on small scales the maxima as well as the minima in the photon transfer function at $\ztrans$ are negative, the heights of the corresponding peaks in $\delta_{\gamma}^2$ are reduced. Therefore the different effects on the CMB of the models with earlier transitions are not the result of modified gravity as such, but rather the period of effectively $\Lambda$CDM evolution following the transition during which the perturbations in $\delta_{\gamma}$ will have undergone oscillations. For example, a mode which has a maximum in $\delta_{\gamma}$ at the transition might have become a minimum by recombination. As the maximum would have been higher compared to the $\Lambda$CDM case so the minimum would be lower and this would generate an enhanced even peak in the $C_l$'s.

\begin{figure}
\begin{center}
\includegraphics[width=12cm]{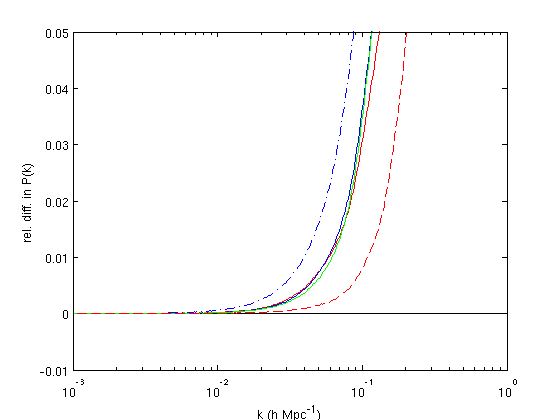}
\end{center}
\caption{Relative differences between the matter power spectra of the modified gravity models and that of a $\Lambda$CDM model. Transition models: $\ztrans = 1090$ and $\beta_i/m_0 = 5\times 10^{6} \Mpc$ (red), $\ztrans = 3000$ and $\beta_i/m_0 = 5\times 10^{7} \Mpc$ (blue), $\ztrans = 5000$ and $\beta_i/m_0 = 1.67\times 10^{8} \Mpc$ (green), $\ztrans = 10^4$ and $\beta_i/m_0 = 5\times 10^{8} \Mpc$ (red dashed) and a generalised chameleon model with $m_0 = 10^{8}, \beta_0 = 9\times 10^{7}, p=3$ and $b=2$ (blue dash-dotted).}\label{fig:Pkrdffs}
\end{figure}

As can be seen in Fig. \ref{fig:Pkrdffs} the effect on the linear matter power spectrum is an increase of power on small scales which is greater for higher values of $\beta_i/m_0$. This is due to the enhanced growth on small scales of the CDM (and to a lesser extent baryon) perturbations before last scattering (see Fig. \ref{fig:transfers}) which occurs despite the fact that, as a result of the very high effective mass of the field, effectively all scales ($k \lesssim 10^{12} \Mpc^{-1}$) are always outside the Compton wavelength. This is because before the transition the couplings to matter are so strong that the factors containing $\beta$'s in Eq. \eqref{eq:deltac''} are still significant. The deviations from $\Lambda$CDM become noticeable at lower $k$ values for later transitions. Roughly speaking, for cases with $\sim 1\%$ deviation in the $C_l$'s, the relative difference in $P(k)$ only becomes much more than $1\%$ at about $k=0.1h\Mpc^{-1}$. One should expect non-linear effects to become important on smaller scales earlier than in the $\Lambda$CDM model, therefore non-linear structures on small scales form earlier in the models we consider in this paper.

It should be noted that for the cases shown, for which $\beta_0 = 1$, the effect of the coupling after the transition is negligible and the observed modifications to the matter power spectrum are entirely due to the anomalous growth before the transition at $\ztrans$ after which the equations governing the evolution of linear perturbations are effectively those of a $\Lambda$CDM model. We also note that reducing the value of $C$ (while keeping the other parameters fixed), and therefore extending the duration of the transition, results in smaller deviations from $\Lambda$CDM, particularly on smaller scales.

Compared to the constraints coming from the local tests and the variation of particle masses, the CMB probes a different part of the parameter space of the model. For values of the coupling $\beta$ of order of unity and masses of the order of 1$\Mpc^{-1}$, there is no effect on the CMB whereas the parameters are excluded by lunar ranging tests. However, if the coupling can take very large values prior to recombination (e.g. $\beta \sim 10^{14}$ for $\ztrans \sim 10^4$) then the CMB constrains the model more strongly than all the local tests.

\subsection{Generalised chameleons}

\begin{figure}
\begin{center}
\includegraphics[width=8cm]{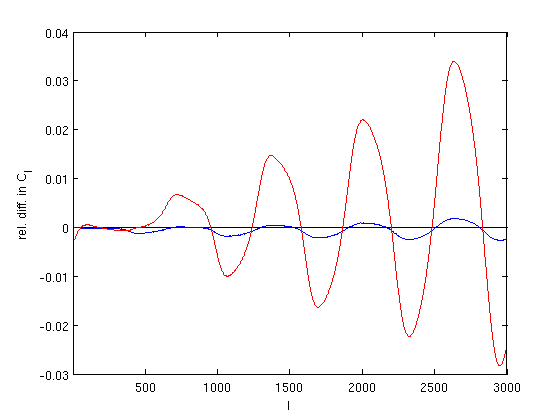}
\includegraphics[width=8cm]{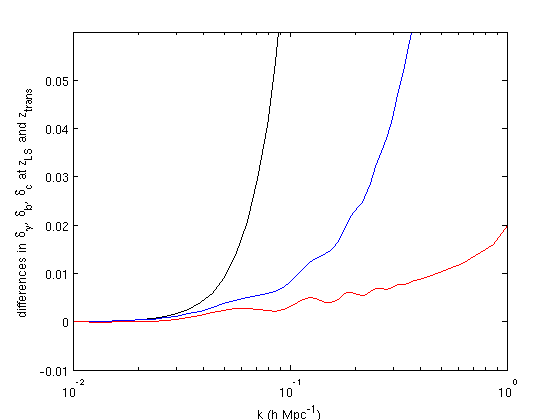}~~~~\includegraphics[width=8cm]{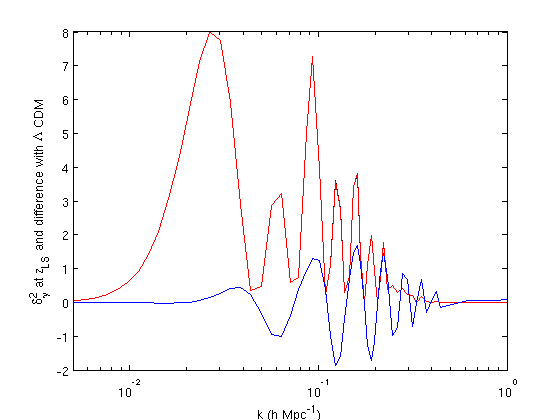}
\end{center}
\caption{Generalised chameleon model with $m_0 = 10^{8}, \beta_0 = 9\times 10^{7}, p=3$ and $b=2$. Top: relative difference between the modified $C_l$'s and those of a $\Lambda$CDM model for the generalised chameleon model (blue) compared with the transition model with $\ztrans = 1090$ (red). Bottom left: differences between the photon (red), baryon (black) and CDM (blue) transfer functions at $z_{LS}$ for the generalised chameleon and those a $\Lambda$CDM model. Bottom right: $\delta_{\gamma}^2$ for the generalised chameleon (red) and a $\Lambda$CDM model (black) at $z_{LS}$ and the magnified difference with $\Lambda$CDM (blue). Note that the chameleon model is  ruled out by BBN constraints and is shown for illustration only.}\label{fig:gencham}
\end{figure}

For the generalised chameleon models we find that for parameters satisfying the constraints given in sections \ref{BBN} and \ref{localtests} there are no visible effects on the CMB but the matter power spectrum is affected in the usual way. In Figs. \ref{fig:Pkrdffs} and \ref{fig:gencham} we can show results for an example case ruled out by the BBN constraint. The enhanced growth in $P(k)$ on small scales is greater than each of the transition examples but the effects on the $C_l$'s are much smaller. However, we also see that as the difference in the photon transfer function is always positive the amplitude of the peaks in $\delta_{\gamma}^2$ is alternately enhanced and reduced and the oscillations in the relative difference of the $C_l$'s are in phase with those of the model with a transition at last scattering. This shows that the effects have the same origin.

In contrast to the transition models, we therefore find that the CMB does not provide complementary, additional constraints on generalised chameleons.

\section{Conclusions}

We have studied models of modified gravity where the screening effects at late time and locally in the solar system still allow for observable effects on cosmological perturbations prior to the recombination era.

We have investigated two types of models: the generalised chameleons with an increasing coupling to matter in the past and a new transition model where the coupling to matter is significantly larger before recombination compared to late times. We have presented analytical estimates and full numerical results using a modified version of CAMB which takes into account the effects of modified gravity. We find that even when the constraints from local tests are satisfied the transition model can produce percent level deviations from $\Lambda$CDM in the CMB angular power spectrum and that these deviations take the characteristic form of alternating enhancements and reductions of power. For the generalised chameleons though it appears that models satisfying the BBN and local test constraints cannot leave observable signatures in the CMB. We should emphasise that local constraints and the BBN constraints are already quite strong constraints. The examples we have considered here predict only small deviations (at most a few percent) of the CMB anisotropies from the $\Lambda$CDM model.

In both the transition and the generalised chameleon models, the effective sound speed of the coupled baryon--photon plasma is the same as it is in the $\Lambda$CDM model to high accuracy. Therefore, the sound horizon is the same. As a consequence, there are no shifts of the acoustic peaks in the CMB or the peaks in the baryonic acoustic oscillations. For the models we considered here, we were not able to find an example for which the sound speed differs significantly from the $\Lambda$CDM value without violating the local or BBN constraints. An additional consequence is that the $\mu$--distortion of the CMB spectrum generated by the dissipation of the acoustic oscillations is the same as in $\Lambda$CDM \cite{vandeBruck:2012vq}.

The effects on the matter power spectrum for both models are the same: an enhancement of power on small length scales (large wave numbers). On the other hand, the predictions for the CMB anisotropy differ between the models. Therefore it may be possible to distinguish different models of modified gravity using the CMB.

\section{Acknowledgements}

PB acknowledges support from the Agence Nationale de la Recherche under contract ANR 2010 BLANC 0413 01. The work of CvdB is supported by the Lancaster-Manchester-Sheffield Consortium for Fundamental Physics under STFC grant ST/J000418/1. ACD is supported in part by STFC. The work of SC has been supported by the Wiener Anspach foundation, the Alexander von Humboldt foundation and the \textit{mandat de retour} program of the Belgian Science Policy (BELSPO). GS is supported by an STFC doctoral fellowship.

\section*{Appendix: Numerical implementation}

CAMB solves the perturbation equations in the synchronous gauge. To account for screened modified gravity it was necessary to add the perturbed Klein Gordon equation which, after using Eq. \ref{eq:m_eff} to write $V_{,\phi\phi}$ in terms of the effective mass $m(a)$, can be written
\begin{equation} \label{eq:pertKGsyn}
\delta\phi'' +2{\cal H} \delta \phi' + (k^2 + a^2 m^2 - a^2 \beta_c^2 \rho^E_c - a^2 \beta_b^2 \rho^E_b )\delta \phi + \tfrac{1}{2}h'\phi' = -a^2 \left(\beta_c \delta\rho_c +\beta_b \delta\rho_b \right),
\end{equation}
and also the evolution equation for $\theta_c$
\begin{equation} \label{eq:thetacsyn}
\theta_c'= -{\cal H} \theta_c + k^2 \beta_c \delta \phi -\beta_c \phi' \theta_c~,
\end{equation}
which, in the presence of modified gravity, is no longer generally vanishing in the synchronous gauge. Extra terms also had to be included in the equations for $\delta_c$, $\delta_b$ and $\theta_b$.

Before last scattering modified gravity also introduces new terms in the calculation of the slip, defined as $\theta_b'-\theta_\gamma' $. In the synchronous gauge the evolution equations for $\thetab$ and $\thetagamma$ are
\begin{equation}  \label{eq:thetabsyn}
\theta_b'= -{\cal H}\theta_b + c_{sb}^2 k^2 \deltab -\frac{an_e \sigma_T}{R} (\theta_b-\theta_\gamma)+ \fMG,
\end{equation}
and
\begin{equation} \label{eq:thetagammasyn}
\theta_\gamma'= \frac{k^2}{4} \delta_\gamma - k^2 \sigmagamma + an_e \sigma_T (\theta_b-\theta_\gamma),
\end{equation}
where $c_{sb}^2$ is the adiabatic sound speed of the baryons (not to be confused with $c_{s}^2$, the sound speed of the coupled baryon-photon fluid) and we have introduced the notation
\be
\fMG \equiv \betab k^2 \deltaphi - \beta_b \phi' \thetab
\ee
corresponding to the additional terms due to screened gravity. During the tight coupling regime the Thomson drag terms in Eqs.~\ref{eq:thetabsyn} and~\ref{eq:thetagammasyn} take very large values making these equations difficult to integrate numerically. In CAMB an alternative form of these equations is used which is valid in the regime $\tau_c \ll \tau$ and $k \tau_c \ll 1$. In the following we follow \cite{MaBert95} to derive the equivalent set of equations in the presence of screened modified gravity. The first step is to use Eq.~\ref{eq:thetagammasyn} to write
\be \label{eq:star1}
(\thetab - \thetagamma) / \tauc = - \thetagamma' + k^2 \left(\frac{1}{4}\deltagamma - \sigmagamma \right),
\ee
and then substitute the corresponding term into Eq.~\ref{eq:thetabsyn}. One gets
\be \label{eq:star2}
\thetab' = - \frac{a'}{a} \thetab + c_{sb}^2 k^2 \deltab - R \thetagamma' + R k^2 \left(\frac 1 4 \deltagamma - \sigmagamma \right) + \fMG~.
\ee
Then one can rewrite $\thetagamma ' = \thetab ' + (\thetagamma' - \thetab') $ in Eq.~\ref{eq:star1}, and replace $(\thetagamma' - \thetab')$ by using Eq.~\ref{eq:star2}. By defining the functions
\be
f \equiv \frac{\tauc}{1+R}
\ee
and
\be
g \equiv - \frac{a'}{a} \thetab + c_{sb}^2 k^2 \deltab - R \thetagamma' + R k^2 (\frac 1 4 \deltagamma - \sigmagamma ) + \fMG~,
\ee
one obtains
\be
\thetab - \thetagamma = f \left[ g - (fg)'  \right] + \mathcal O (\tauc^3)~,
\ee
in which the $\tauc^3$ terms can be conveniently neglected. Differentiating this equation gives
\be \label{eq:star3}
\thetab' - \thetagamma' = \frac{f'}{f} (\thetab - \thetagamma) + f (g'  f'' g - 2 f' g' - g'' f)~,
\ee
with
\be
g' = - \mathcal H \thetab' - \mathcal H' \thetab + k^2 \left[ (c_{sb}^2)' \deltab + c_{sb}^2 \deltab' - \frac 1 4 \deltagamma' + \sigmagamma' \right] + \fMG'~.
\ee
The next step involves the following trick. First, one can add $- \mathcal H \thetab' + \mathcal H \thetab'  $ to the right hand side of the last equation. Then Eq.~\ref{eq:thetabsyn} can be used to express the $+ \mathcal H \thetab'$ term. Finally one can rewrite $- \mathcal H \thetab' = - \mathcal H (\thetab' - \thetagamma') - \mathcal H \thetagamma'$ and use Eq.~\ref{eq:thetagammasyn} to express the last term.
After using ${c_{sb}^2}' = - c_{sb}^2 \mathcal H$, one can obtain
\be
g' = -2 \mathcal H (\thetab' - \thetagamma' ) - \frac{a''}{a} \thetab + k^2 \left[ - \frac 1 2 \mathcal H \deltagamma
+ 2 \mathcal H \sigmagamma + c_{sb}^2 \deltab' - \frac 1 4 \deltagamma' + \sigmagamma' \right]
+ \fMG' + \left(\frac{\mathcal H R+ 2 \mathcal H}{\tauc}   \right) (\thetagamma - \thetab)~.
\ee
Finally, keeping only the terms in  $\mathcal O(\tau_c)$, Eq.~\ref{eq:star3} reads
\be \label{eq:star4}
\thetab ' - \thetagamma' = \left( \frac{\tauc'}{\tauc} - \frac{2 \mathcal H}{1+R} \right) (\thetab - \thetagamma)
+ \frac{\tauc}{1+R} \left[- \frac{a''}{a} \thetab - \frac 1 2 \frac{a'}{a} k^2 \deltagamma + k^2 (c_{sb}^2 \deltab' - \frac 1 4 \deltagamma)
+ \fMG' + \mathcal H \fMG \right]
\ee
During the tight coupling regime $\thetab$ is obtained by integrating
\be \label{eq:star5}
\thetab ' = \frac{1}{1+R} \left[ - \frac{a'}{a} \thetab + c_{sb}^2 k^2 \deltab + k^2 R (\frac 1 4 \deltagamma - \sigmagamma) + \fMG  \right]
+ \frac{R}{1+R} (\thetab' - \thetagamma')~,
\ee
which is derived directly from the Eq.~\ref{eq:star2} and in which the slip is given by Eq~\ref{eq:star4}.

Compared to the standard general relativistic case we get two contributions from modified gravity. The first is the term $\fMG$ in Eq.~\ref{eq:star5} and the second comes from an additional $\tauc (\fMG'+\mathcal H \fMG)/(1+R)$ term in the slip.

The derivative of $\fMG$ is given by
\ba	
\fMG' & = & \betab' k \deltaphi + \betab k \deltaphi' - \betab' \phi' \thetab - \betab \phi'' \thetab - \betab \phi' \thetab' \\
      & = & \betab' k \deltaphi + \betab k \deltaphi' - \betab' \phi' \thetab - \betab \phi'' \thetab - \betab \phi' (- \frac{a'}{a} \thetab + c_{sb}^2 k^2 \deltab + \fMG ) + \frac{R}{\tauc} \betab \phi' (\thetab - \thetagamma)~,
\ea
where the last equation is obtained after using Eq.~\ref{eq:thetabsyn}.

To get an equation for $\thetagamma'$ we can use Eq.~\ref{eq:thetabsyn} to express the drag term as
\be
(\thetab - \thetagamma) / \tauc = -R(\thetab' + {\cal H}\theta_b - c_{sb}^2 k^2 \deltab - \fMG)
\ee
and then substitute this in Eq.~\ref{eq:thetagammasyn} to obtain
\be
\theta_\gamma'= \frac{k^2}{4} \delta_\gamma - k^2 \sigmagamma - R(\thetab' + {\cal H}\theta_b - c_{sb}^2 k^2 \deltab - \fMG).
\ee
This last equation is used at all times in CAMB with $\thetab'$ determined by Eq.~\ref{eq:star5} during the tight-coupling regime and by Eq.~\ref{eq:thetabsyn} otherwise.

We have also modified the source terms. These are computed by integrating by parts the line of sight integral giving the CMB temperature multipoles today. This is required in CAMB for the Bessel functions to be the only $l$-dependent variables in the remaining line of sight integral (see Seljak and Zaldarriaga for more details). However, we find that modifying the source terms does not have any visible effects on the CMB angular power spectrum for the models and parameters we have considered.

Finally, in order to avoid the time-consuming numerical integration of the field perturbations at early times when they oscillate quickly, we have introduced the approximation
\be
\delta \phi = -\frac{\beta_c \delta\rho_c +\beta_b \delta\rho_b } {\frac{k^2}{a^2}+m^2-\beta_c^2 \rho^E_c -\beta_b^2 \rho^E_b }
\ee
when the condition
\be
(k^2 + a^2 m^2 - a^2 \beta_c^2 \rho^E_c - a^2 \beta_b^2 \rho^E_b)\delta\phi \gg |2 \mathcal H \delta \phi' + \tfrac{1}{2}h'\phi'|
\ee
is satisfied.

\end{document}